%
\input harvmac
\def\Title#1#2#3{#3\hfill\break \vskip -0.35in
\rightline{#1}\ifx\answ\bigans\nopagenumbers\pageno0\vskip.2in
\else\pageno1\vskip.2in\fi \centerline{\titlefont #2}\vskip .1in}


\def\R{\hbox{\rm I \kern-5pt R}}

\font\ticp=cmcsc10
\def\ajou#1&#2(#3){\ \sl#1\bf#2\rm(19#3)}
%
%
\lref\akprl{A.~Kent, \ajou Phys. Rev. Lett. &78 (97) 2874.}
 
\lref\akordered{A.~Kent, ``Quantum Histories and Their Implications'',
gr-qc/9607073, submitted to Ann. Phys.}

\lref\dowkerkentone{F.~Dowker and A.~Kent, 
\ajou J. Stat. Phys. &82 (96) 1575.}

\lref\dowkerkenttwo{F.~Dowker and A.~Kent, \ajou Phys. Rev. Lett. &75
(95) 3038.}

\lref\comment{R.~Griffiths and J.~Hartle, gr-qc/9710025, to appear in
Phys. Rev. Lett.}

\lref\kenttwo{A.~Kent, \ajou Phys. Rev. A &54 (96) 4670.}

\lref\akscripta{A.~Kent, ``Quantum Histories'', Proceedings of the 
104th Nobel Symposium, Gimo, June 1997, to appear in Physica
Scripta.} 

\Title{\vbox{\baselineskip12pt\hbox{ DAMTP/97-126}\hbox{ gr-qc/9808016}{}}
}{{\centerline{\vbox{\hbox{Consistent Sets and Contrary Inferences:}
\hbox{\qquad Reply to Griffiths and Hartle}}}}}{~}

\centerline{{\ticp Adrian Kent\foot{Email: apak@damtp.cam.ac.uk}}}
\vskip.1in
\centerline{\sl Department of Applied Mathematics and
Theoretical Physics,}
\centerline{\sl University of Cambridge,}
\centerline{\sl Silver Street, Cambridge CB3 9EW, U.K.}

\bigskip
\centerline{\bf Abstract}
{It was pointed out recently [A. Kent, Phys. Rev. Lett. 78 (1997)
2874] that the consistent histories
approach allows contrary inferences to be made from the same
data.  These inferences correspond to projections $P$ and
$Q$, belonging to different consistent sets, with the properties 
that $PQ = QP = 0$ and $P \neq 1 -Q$.  To many, this seems undesirable 
in a theory of physical inferences.  It also raises a specific 
problem for the consistent histories formalism, since that 
formalism is set up so as to eliminate contradictory inferences,
i.e. inferences $P$ and $Q$ where $P = 1 - Q$.  
Yet there seems to be no sensible physical distinction between contradictory
and contrary inferences.  

It seems particularly hard to defend the 
asymmetry, since (i) there is a well-defined quantum
histories formalisms which admits both contradictory and 
contrary inferences, and (ii) there is also a well-defined formalism, 
based on ordered consistent sets of histories, which 
excludes both.   

In a recent comment, Griffiths and Hartle, while accepting the
validity of the examples given in the above paper, restate their own
preference for the consistent histories formalism. 
As this brief reply explains, in so doing, they fail to address the 
arguments made against their approach to quantum theory.
\vfill
To appear in Phys. Rev. Lett.}
\vfill
\eject
Griffiths and Hartle\refs{\comment} 
correctly reiterate that logical contradictions in the consistent
histories formalism can be avoided by restricting one's reasoning
to a single consistent set.   
This is, of course, generally understood, and clearly stated
in the original Letter.\refs{\akprl} 
But it is no great virtue.  {\it Any} formalism which 
ascribes probability distributions to sets of histories in quantum 
theory can be made free from contradiction by such a restriction, 
whether or not any consistency criterion is 
imposed on the sets, whatever rule is used to define history probabilities.  
There are infinitely many such formalisms,  and {\it a priori} the 
consistent histories formalism has no special status.  
The questions we should ask of it, or any other quantum 
history formalism, are whether its rules 
are natural and lead to physically plausible and scientifically
useful conclusions.  It is the naturality of the consistency criterion
and the plausibility of its physical implications which the Letter
addresses.  (Criticisms of the consistent histories approach
as a scientific theory can be found
elsewhere.\refs{e.g. \dowkerkentone, 
\dowkerkenttwo, \kenttwo, \akscripta})

To examine any history-based formalism properly, it 
is necessary to stand outside that formalism, to ask: 
Why these rules, rather than others?  What are their
physical consequences?  How do they compare with those
of alternative rules?  To attempt to defend the consistent 
histories formalism simply by restating the internal consistency of
its own rules, and to argue that a particular asymmetry is acceptable
simply because it is a feature of the formalism, as Griffiths and
Hartle in effect do, is to fail to address these key questions.

Our understanding of what it means for two physical propositions to be 
contradictory or contrary is --- historically and, unless we are 
already irrevocably committed to the formalism under discussion, 
logically --- prior to our understanding of the features of 
a given formalism.  Our views on whether the formalism is 
natural, plausible or useful are, inevitably, framed in 
terms of that prior understanding.  
The standard contradictory/contrary terminology thus seems to me 
most appropriate, though of course the argument is invariant 
under translation.  

To recap: the Letter points out that
the consistent histories formalism allows contrary inferences, although
one of the arguments used in its justification is 
that it prevents contradictory inferences.  
As Ref. \refs{\comment} notes, ``There are no initial and final
states for which projectors $P$ and $(1-P)$ can have probability
one in different consistent families, because the probability 
assigned to any event on the basis of given data is independent
of the consistent family''.  
But --- it is vital to be clear on this --- this is a 
deliberately chosen feature, not a logically necessary requirement. 
In many history-based formalisms of quantum theory --- for example, 
that based on unrestricted sets of histories --- a proposition 
$P$ can have probability one in one set and its complement 
$(1-P)$ non-zero probability in another, conditioned 
on the same data.  These formalisms too 
produce no inconsistency if reasoning is carried out only 
within one set.\refs{\dowkerkenttwo, \akscripta}  
The term ``consistent histories'' is in this 
respect misleading. 

No argument for the asymmetry between contradictory and contrary
inferences has yet been produced. To say\refs{comment} ``a pair $P$ and $1-P$ 
can always be associated with the logical notion of `contradictory', 
while perpendicular projectors need not be associated with the logical
notion of `contrary' '' is, again, to report a feature of the
consistent histories formalism, not a general truth.
In attempting to defend the asymmetry, Griffiths and Hartle simply
restate it.  
 
Consistent historians must also accept 
that contrary statements about past events --- ``the electron went
through slit A with probability one'' and ``the electron went 
through slit B with probability one'', say ---
can be equally valid, and hence that standard ordering 
inferences based on our observations can fail. 
For example, the statement that a particle 
was observed in one region need not imply that it was observed
in a region containing the first, even when each  
statement can consistently be considered together with our 
observations.\refs{\akordered} 

I should stress again that these facts do not amount to 
a logical refutation of the consistent histories approach.  
Readers will no doubt form their own views of their implications 
for its naturality and plausibility.  
\vskip5pt
\leftline{\bf Acknowledgements}  
I am very grateful to Fay Dowker and Arthur Fine for valuable 
discussions, to Bob Griffiths and Jim Hartle for 
helpful comments, and to the Royal Society for financial support. 
\listrefs
\end